\documentclass[aps, prb,reprint,superscriptaddress,showpacs]{revtex4-1}

\usepackage{graphicx}
\usepackage{dcolumn}
\usepackage{bm}
\usepackage{natbib}
\usepackage{amsfonts}
\usepackage{amsmath}
\usepackage{amssymb}

\begin{document}

\title{Direct observation of charge order in underdoped and optimally doped Bi$_{2}$(Sr,La)$_{2}$CuO$_{6+\delta}$ by resonant inelastic x-ray scattering}

\author{Y.~Y.~Peng}
\email{yingying.peng@polimi.it} \affiliation{Dipartimento di Fisica, Politecnico di Milano, Piazza
Leonardo da Vinci 32, I-20133 Milano, Italy}
\author{M. ~Salluzzo}
\affiliation{CNR-SPIN, Complesso MonteSantangelo - Via Cinthia, I-80126
Napoli, Italy}
\author{X.~Sun}
\affiliation{Beijing National Laboratory for Condensed Matter Physics,
Institute of Physics, Chinese Academy of Sciences, Beijing 100190, China}
\author{A.~Ponti}
\affiliation{Istituto di Scienze e Tecnologie Molecolari, Via Camillo
Golgi 19, I-20133 Milano, ITALY}
\author{D.~Betto}
\affiliation{ESRF, The European Synchrotron, CS 40220, F-38043 Grenoble
Cedex, France}
\author{A.~M.~Ferretti}
\affiliation{Istituto di Scienze e Tecnologie Molecolari, Via Camillo
Golgi 19, I-20133 Milano, Italy}
\author{F.~Fumagalli}
\affiliation{Italian Institute of Technology-Center for Nanoscience and
Technology, Via Pascoli, 70/3, 20133 Milano Italy}
\author{K.~Kummer}
\affiliation{ESRF, The European Synchrotron, CS 40220, F-38043 Grenoble
Cedex, France}
\author{M.~Le Tacon}
\affiliation{Max-Planck-Institut f\"{u}r Festk\"{o}rperforschung,
Heisenbergstra{\ss}e 1, D-70569 Stuttgart, Germany}
\affiliation{Karlsruher Institut f\"{u}r Technologie Institut f\"{u}r
Festk\"{o}rperphysik Hermann-v.-Helmholtz-Platz 1, D-76344
Eggenstein-Leopoldshafen, Germany}
\author{X.~J.~Zhou}
\affiliation{Beijing National Laboratory for Condensed Matter Physics,
Institute of Physics, Chinese Academy of Sciences, Beijing 100190, China}
\author{N.~B.~Brookes}
\affiliation{ESRF, The European Synchrotron, CS 40220, F-38043 Grenoble
Cedex, France}
\author{L.~Braicovich}
\affiliation{Dipartimento di Fisica, Politecnico di Milano, Piazza
Leonardo da Vinci 32, I-20133 Milano, Italy} \affiliation{CNR-SPIN and CNISM,
Politecnico di Milano, Piazza Leonardo da Vinci 32, I-20133 Milano, Italy}
\author{G.~Ghiringhelli}
\email{giacomo.ghiringhelli@polimi.it} \affiliation{Dipartimento di Fisica, Politecnico di Milano, Piazza
Leonardo da Vinci 32, I-20133 Milano, Italy} \affiliation{CNR-SPIN and CNISM,
Politecnico di Milano, Piazza Leonardo da Vinci 32, I-20133 Milano, Italy}

\date{\today}

\begin{abstract}

Charge order in underdoped and optimally doped high-$T_\mathrm{c}$
superconductors Bi$_{2}$Sr$_{2-x}$La$_x$CuO$_{6+\delta}$ (Bi2201) is
investigated by Cu $L_3$ edge resonant inelastic x-ray scattering (RIXS).
We have directly observed charge density modulation in the optimally doped
Bi2201 at momentum transfer $Q_{\|} \simeq 0.23$ rlu, with smaller intensity and correlation length with
respect to the underdoped sample. This demonstrates the short-range charge
order in Bi2201 persists up to optimal doping, as in other hole-doped
cuprates. We explored the nodal (diagonal) direction and found no charge
order peak, confirming that charge order modulates only along the Cu-O
bond directions. We measured the out-of-plane dependence of charge order,
finding a flat response and no maxima at half integer \emph{L} values.
This suggests there is no out-of-plane phase correlation in single layer
Bi2201, at variance from YBa$_2$Cu$_3$O$_{6+x}$ and
La$_{2-x}$(Ba,Sr)$_x$CuO$_4$. Combining our results with data from the
literature we assess that charge order in Bi2201 exists in a large doping
range across the phase diagram, i.e. $0.07 \lesssim p \lesssim 0.16$,
demonstrating thereby that it is intimately entangled with the
antiferromagnetic background, the pseudogap and superconductivity.
\end{abstract}

\pacs{74.72.Gh,74.25.Jb,74.25.Gz,74.72.Kf}

\maketitle

\section{Introduction}

The search for the underlying mechanism of high-$T_\mathrm{c}$
superconductivity in cuprates remains active for three decades after its
discovery \cite{HighTcuprate}. The insulating parent compounds
become superconductors by chemical doping, which modifies the charge
balance of the CuO$_2$ planes and rapidly suppresses their 2D long-range
antiferromagnetic order \cite{PALeeReview}. In the ``normal"
state, above the superconducting critical temperature $T_\mathrm{c}$,
there is an exotic pseudogap phase whose origin and relation to the
superconducting phase are still much debated \cite{PGreview}. More
recently, evidence of charge order, or charge density wave (CDW), within
the CuO$_2$ planes has been found, below optimal doping, in several
families,
\cite{TranquadaStripe,FujitaStripe,JapaneseLBCO,Stripexray,CDWLBCOPRB,JulienNMR,GiacomoCDW,RXSCDW,changCDW,diffractionCDW,Damascelliscience,STMBi2212,HashimotoBi2212,HBCOCDW,TaconYBCOPRB,CominNBCO,EduardoSA}
further increasing the complexity of the cuprates' phase diagram
\cite{KeimerNature}. The temperature evolution of CDW
\cite{GiacomoCDW} and its behavior under magnetic fields
\cite{changCDW} have indicated that charge order is in competition with
superconductivity. Although the phenomenology of CDW has grown fast, it is
still patchy and a systematic knowledge of its doping evolution would help
clarify its role in high-$T_\mathrm{c}$ superconductivity and its relation
with the quantum critical points (QCP) in the phase diagram.

Early evidence of bulk charge order was obtained in La-based
cuprates by inelastic neutron scattering (INS) near the hole doping $p =
^{}1\!/_8$ [\onlinecite{TranquadaStripe},\onlinecite{FujitaStripe}] and
later by x-ray scattering with an approximately commensurate wave-vector
$Q_\| \simeq 0.25$ reciprocal
lattice units (rlu) [\onlinecite{Stripexray}]. More recently, an
incommensurate charge order at $Q_\| \simeq 0.31$ rlu along the Cu-O bond
direction, has been observed by various techniques in
(Y,Nd)Ba$_2$Cu$_3$O$_{6+x}$ (YBCO,
NBCO)\cite{JulienNMR,GiacomoCDW,RXSCDW,changCDW,diffractionCDW},
which might be responsible for the Fermi surface reconstructions in high
magnetic fields giving rise to quantum oscillations
\cite{quantumoscillation1,quantumoscillation2}. Soon
after, the enhanced sensitivity of resonant x-ray scattering has also
allowed the detection of a short-range charge order in
Bi$_2$Sr$_{2-x}$La$_x$CuO$_{6+\delta}$ (Bi2201)
[\onlinecite{Damascelliscience}], Bi$_2$Sr$_2$CaCu$_2$O$_{8+\delta}$
(Bi2212) [\onlinecite{STMBi2212},\onlinecite{HashimotoBi2212}] and
HgBa$_2$CuO$_{4+\delta}$ [\onlinecite{HBCOCDW}], and eventually in
electron-doped (Nd,La)$_{2-x}$Ce$_x$CuO$_4$
[\onlinecite{CominNBCO},\onlinecite{EduardoSA}], indicating its ubiquity
in cuprate superconductors. Hereafter we will confine our discussion to
hole-doped systems for brevity. The CDW is strongest in the underdoped
regime and persists up to optimal doping
\cite{CDWLBCOPRB,HashimotoBi2212,TaconYBCOPRB}.
In Bi2201, the charge order was observed by resonant x-ray scattering
(RXS) in the underdoped region ($p\sim 0.115 - 0.145$) with wave vector, decreasing with $p$,
that was proposed to match the distance between the tips of the
ungapped segments of Fermi surface (``Fermi arcs'')
\cite{Damascelliscience}. A previous resonant inelastic x-ray
scattering (RIXS) study on optimally doped Bi2201 (OP-Bi2201) has not
found a charge order signal directly \cite{PengPRB}. Instead a
low energy feature at $Q_\| \simeq 0.22$ rlu was observed up to 200K and
was attributed to a phonon signal. On the other hand, in OP-Bi2201
angle-resolved photoemission spectroscopy (ARPES) has shown a
particle-hole symmetry breaking and a phase transition below the pseudogap
temperature \cite{Hashimotonphys,Hescience}. And
scanning tunneling microscopy (STM) has found a checkerboard-like
electronic modulation in a broad doping range of Bi2201
[\onlinecite{WDWise}]. However, electronic states may vary in bulk
(studied with x-rays) and at the surface (studied with STM and ARPES).

The discrepancy between bulk and surface measurements calls for a more
accurate investigation: here, by using high resolution RIXS at the Cu $L_3$
edge, we study the charge order in underdoped ($T_\mathrm{c} = 15$K,
UD15K) and optimally doped ($T_\mathrm{c} = 33$K, OP33K)
Bi$_{2}$Sr$_{2-x}$La$_x$CuO$_{6+\delta}$. We will focus on the
quasi-elastic spectral component that is sensitive to charge modulations
\cite{GiacomoCDW}. Along the Cu-O bond direction, we observed a bulk charge
order peak at incommensurate vector $Q_{\|} \simeq 0.26$ (0.23) rlu in
UD15K (OP33K). This expands up to optimal doping the region where charge
order, superconductivity and pseudogap coexist in B2201. We also performed
temperature measurements on OP33K across $T_\mathrm{c}$ and $T^*$ to
investigate the relations between charge order, superconductivity and
pseudogap. We noticed that a prior energy-integrated RXS measurement on
Bi2201 reported no CDW signatures along the diagonal (nodal) direction
\cite{nodalCDW}. However, the checkerboard-like features observed by STM
[\onlinecite{WDWise}] are compatible with a two-dimensional (2D) CDW structure, as
opposed to the one-dimensional (1D) stripe-like shape proposed for YBCO
[\onlinecite{nodalCDW}]. Here we use energy-resolved RIXS and its higher
sensitivity to ascertain this issue. Finally, in UD15K we measured the
charge order peak intensity along the $c^*$-direction [we quote
(\emph{H,K,L}) for wave vector coordinates in pseudo-tetragonal structure]
to understand the out-of-plane phase correlation, and we compare the
results to the anti-phase correlations in YBa$_2$Cu$_3$O$_{6+x}$
[\onlinecite{CDWantiphase}] and La$_{2-x}$(Ba,Sr)$_x$CuO$_4$
[\onlinecite{Stripexray,CDWLBCOPRB,JapaneseLBCO}].

\section{Experimental method}

\begin{figure}[tbp]
\begin{center}
\includegraphics[width=0.95\linewidth,angle=0]{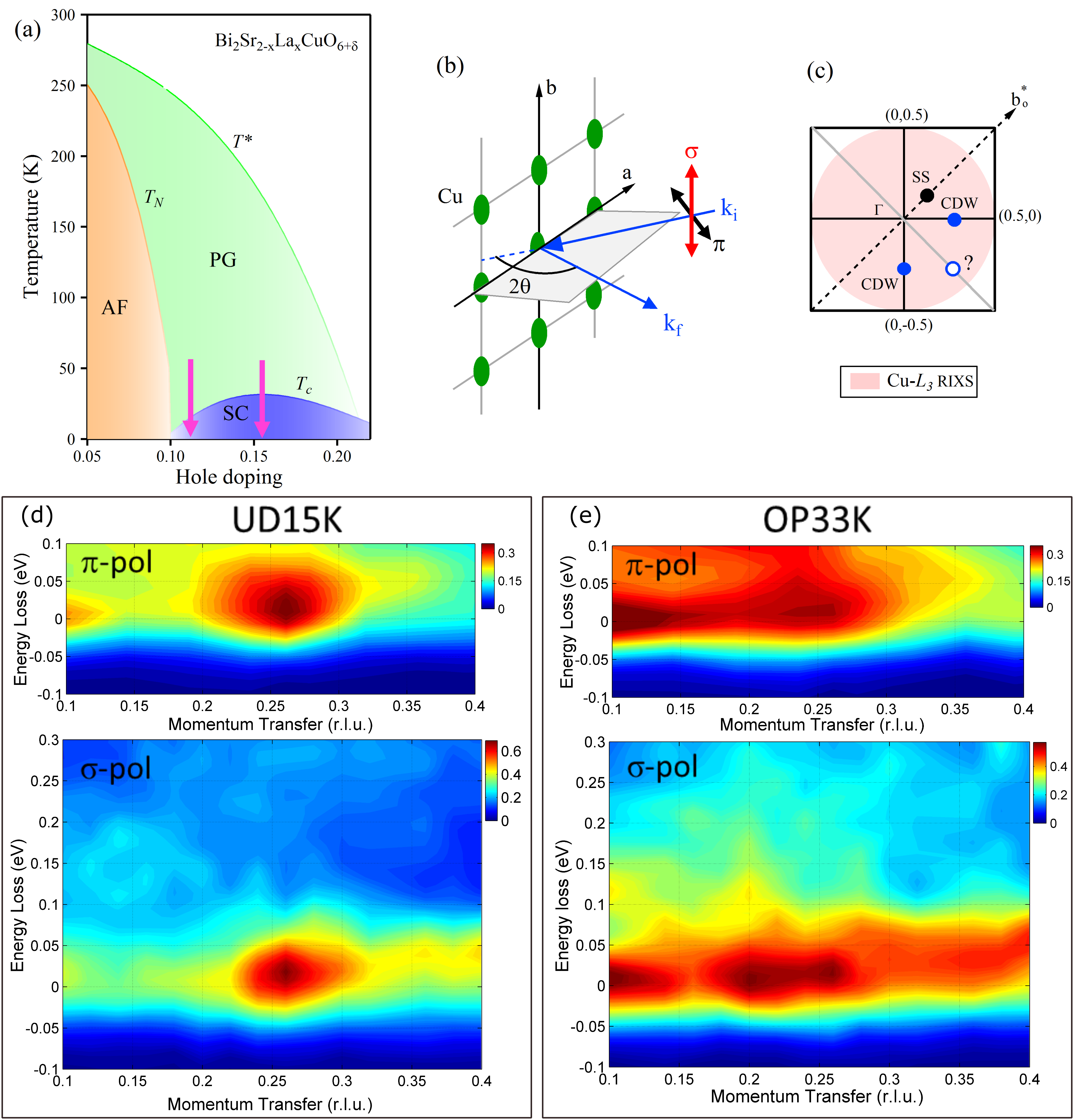}
\end{center}
\caption{(color online) (a) The phase diagram of
Bi$_{2}$(Sr,La)$_{2}$CuO$_{6+\delta}$ [\onlinecite{NMRBi2201}]. It shows
the antiferromagnetic (AF) region defined by $T_N$, superconducting (SC)
region defined by $T_\mathrm{c}$ and the pseudogap (PG) region defined by
$T^*$. Here we study two doping levels as indicated by the arrows. (b) The
experimental geometry. The incident photon polarization can be chosen
parallel ($\pi$) or perpendicular ($\sigma$) to the scattering plane. (c)
Reciprocal-space image, the accessible reciprocal space in Cu $L_3$ RIXS
experiments with 150 deg scattering angle is indicated by the pink circle.
In Bi-based cuprates there is a well-known superstructure (SS) as indicated by
the black circle along the $b_o^*$ direction. (d) Energy/momentum
intensity maps of RIXS spectra along (0,0)-(0.5,0) symmetry direction
taken with $\pi$ or $\sigma$-polarized incident light at 20~K for UD15K.
(e) Same as (d) but for OP33K.}
\end{figure}

The phase diagram of Bi$_{2}$Sr$_{2-x}$La$_x$CuO$_{6+\delta}$
[\onlinecite{NMRBi2201}] is shown in Fig.~1(a). By substitution of Sr by La
we can obtain a wide range of doping. Here we study the underdoped (UD15K,
\emph{p} $\simeq$ 0.115) and the optimally-doped (OP33K, \emph{p} $\simeq$
0.16) samples, as indicated by the arrows in the figure. The high quality
single crystals were grown by the floating zone method. The hole
concentration was optimized by annealing the samples in O$_2$ flow. The
sample growth and characterization methods were reported in Ref.~\onlinecite{meng}. The RIXS measurements were performed with the ERIXS
spectrometer at the beam line ID32 of ESRF (The European
Synchrotron,Grenoble, France). The x-ray energy was tuned to the maximum
of the Cu $L_3$ absorption peak around 931 eV. The experimental energy
resolution was $\sim70$ meV. The samples were cleaved out-of-vacuum just
before installation inside the vacuum measurement chamber, to reduce
surface contamination and oxygen depletion.

The experimental geometry is shown in Fig.~1(b). X-rays are incident on
the sample surface and scattered by $2\theta$, which can be changed
continuously from 50 deg to 150 deg. The x-ray polarization can be chosen
parallel ($\pi$) or perpendicular ($\sigma$) to the horizontal scattering
plane. Reciprocal lattice units (rlu) were defined using the
pseudo-tetragonal unit cell with $a=b=3.86$ {\AA} and $c=24.4$ {\AA},
where the axis ${c}$ is normal to the cleaved sample surface. The sample
can be rotated azimuthally around the ${c}$-axis to choose the in-plane
wave-vector component. We determined accurately the orientations of our
Bi2201 samples by utilizing the [002] Bragg peak and the superstructure
peak. The typical size of the Brillouin zone along [1,0] direction in
cuprates is $\sim 0.81$ {\AA}$^{-1}$ and the maximum total momentum
transfer at the Cu $L_3$ edge with $2\theta=150$ deg is $\sim 0.85$ {\AA}$^{-1}$, 
which allows one to cover the whole first magnetic Brillouin zone as
indicated by the pink area in Fig.~1(c). The well-known incommensurate
supermodulation (superstructure) in the Bi-based cuprates, due to the
distortions of the BiO bi-layers, projects along the $b_o^*$ direction
in the orthorhombic notation \cite{ZChen}, giving a peak
around [$Q_{\mathrm{ss}}, Q_{\mathrm{ss}}$] in pseudo-tetragonal notation,
with $Q_{\mathrm{ss}} \simeq ^{}1\!/_8$ rlu. In the same notation charge order is observed
both along (0,0)-(0.5,0) and (0,0)-(0,0.5) directions. Along the diagonal
we performed the measurement only along (0,0)-(0.5,-0.5) direction, to
avoid confusion with the intense superstructure peak and its replicas along the same direction. We present RIXS
spectra normalized to the integrated intensity of the ${dd}$ excitations
following previous conventions \cite{Luciomagnon}. The zero
energy-loss position was determined by measuring, for each transferred
momentum, a non-resonant spectrum of silver paint or carbon-tape.

\section{Results}
\subsection{Doping dependence}

Figure~1(d) displays the energy/momentum intensity maps of RIXS spectra
for UD15K along (0,0)-(0.5,0) symmetry direction, collected at $T=20$K
with both $\pi$ and $\sigma$ polarized incident x-rays. Both maps exhibit
a charge order signal around the quasi-elastic energy region, with similar
wave vector as reported by RXS [\onlinecite{Damascelliscience}]. In Fig.~1(e), we
have identified a bulk charge order in OP33K with both $\pi$ and $\sigma$
polarized incident x-rays. The charge order signal in OP33K looks much
broader and weaker than that in UD15K, which might have hindered its
discovery in previous studies \cite{PengPRB}. The strong intensity at
small momentum transfer is due to the tails of the elastic peaks arising
from the reflectivity of the surface at specular angle.

\begin{figure}[tbp]
\begin{center}
\includegraphics[width=0.95\columnwidth,angle=0]{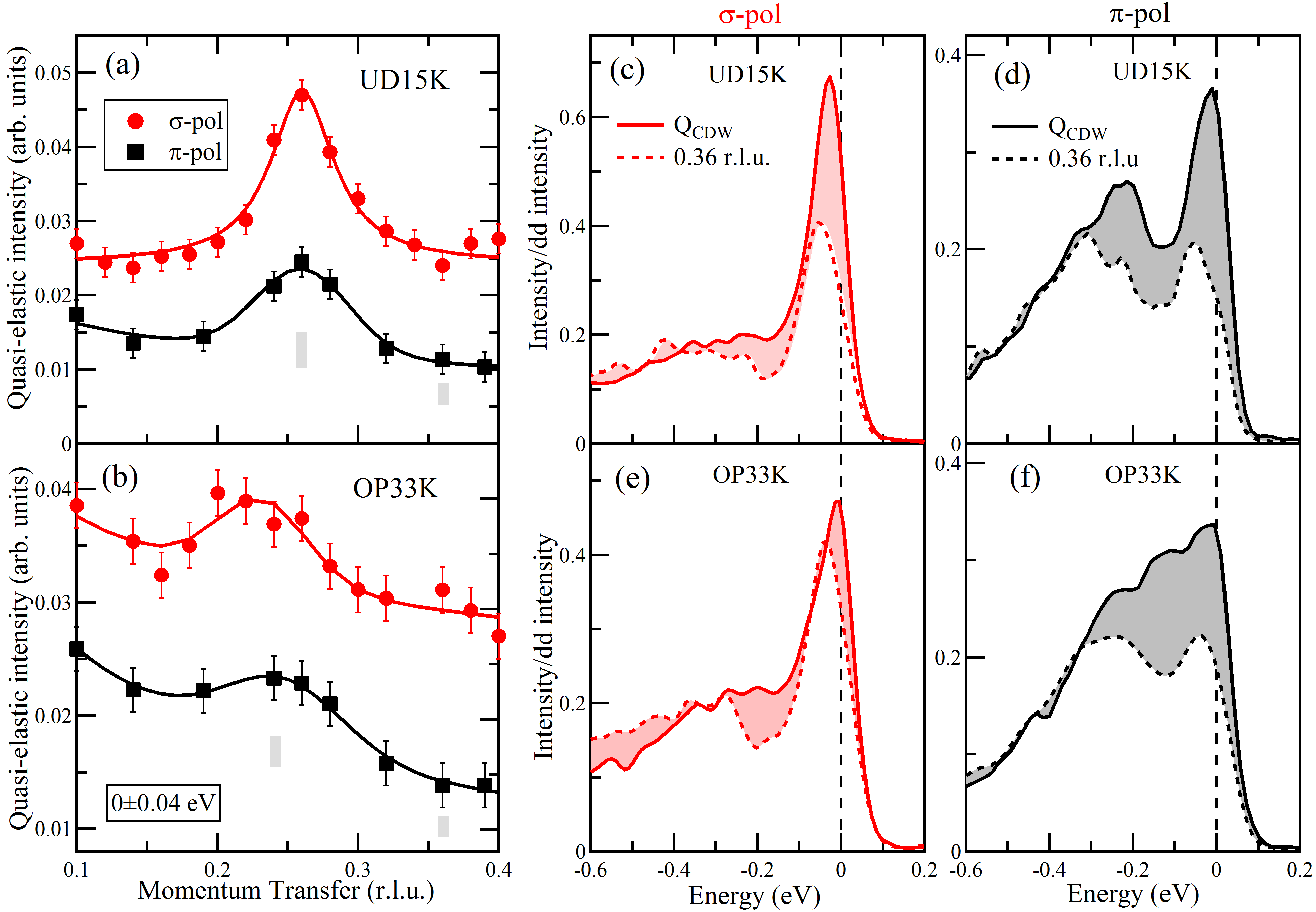}
\end{center}
\caption{(color online) Intensity at 0$\pm$0.04 eV for the quasi-elastic
signal along (0,0)-(0.5,0) symmetry direction with $\pi$- or
$\sigma$-polarized incident light at 20~K for UD15K (a) and OP33K (b).
Solid lines are Lorentzian peak fits to the data with a power law
background. (c,d) Comparing the RIXS spectra of UD15K at
$Q_{\mathrm{CDW}}$ and $Q_{\|}$$\simeq$0.36 rlu as indicated by the gray
bars in (a) with $\sigma$- and $\pi$-polarization respectively. The
differences between the two spectra are highlighted by the red (gray)
shading for $\sigma$ ($\pi$)-polarization. (e,f) Similar to (c,d) but for
OP33K. }
\end{figure}

To better visualize the charge order, we show the integral quasi-elastic
intensities at 0$\pm$0.04 eV for UD15K and OP33K in Fig.~2(a) and (b),
respectively. We can directly observe the charge order with $\sigma$- or
$\pi$-polarization at similar wave vector. We determined the
full-width-half-maximum (FWHM) of charge order from $\sigma$-polarization
with more data points. By fitting the intensity with a power law profile
for the background and a Lorentzian function for the CDW peak, we
determined the charge order vector $\sim 0.26$ rlu for UD15K and $\sim$
0.23 rlu for OP33K. The CDW peak intensity is weaker on top of a higher
background in OP33K. For UD15K we compare two spectra, one at
$Q_{\mathrm{CDW}} \simeq 0.26$ rlu and the other at $Q_{\|} \simeq 0.36$
rlu, outside the CDW region, in Fig.~2(c) and (d) for $\sigma$- and
$\pi$-polarization, respectively. Clearly, the elastic peak is much
stronger at $Q_{\mathrm{CDW}}$. In contrast, the quasi-elastic peak shifts
to higher energy loss at $Q_{\|} \simeq 0.36$ rlu due to the phonon
contributions. In $\pi$-polarization there is also prominent paramagnon
feature, which disperses to higher energy with increasing momentum
transfer as discussed previously \cite{PengPRB}. For OP33K we observe
similar trends in Fig.~2(e) and (f), but the spectral difference between
the two momenta is smaller than in UD15K.

\begin{figure}[tbp]
\begin{center}
\includegraphics[width=0.9\linewidth,angle=0]{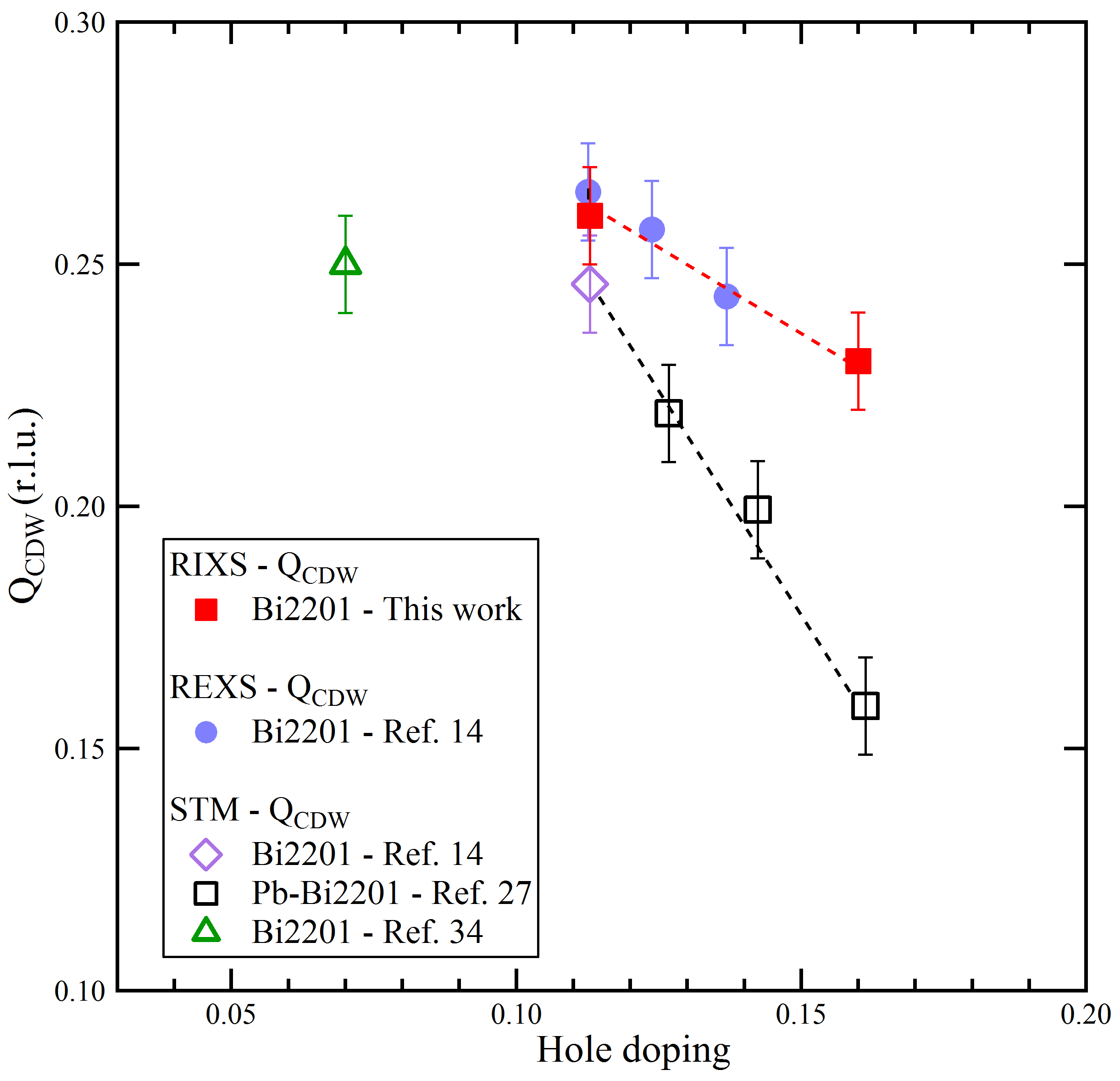}
\end{center}
\caption{(color online) Doping dependence of the charge order wave vector
$Q_{\mathrm{CDW}}$. Data from RXS [\onlinecite{Damascelliscience}] and STM
[\onlinecite{Damascelliscience},\onlinecite{WDWise},\onlinecite{CaiSTM}]
on Bi2201 are included. Bars represent errors due to uncertainty. Dashed
lines are guide for the eye. }
\end{figure}

In Fig.~3, we summarize the doping dependence of the charge order vector
in Bi2201 from measurements by RIXS (our work), RXS
[\onlinecite{Damascelliscience}] and STM
[\onlinecite{Damascelliscience},\onlinecite{WDWise},\onlinecite{CaiSTM}].
The charge order vectors determined with RIXS follow the trend determined
by RXS, while they are significantly larger than those obtained with STM
measurements (see discussion below). When doping changes from \emph{p}
$\simeq$ 0.115 to \emph{p} $\simeq$ 0.16, the FWHM of CDW peak grows from
0.054 to 0.07 rlu, indicating a decreasing coherence length.

\subsection{Momentum dependence}

\begin{figure}[tbp]
\begin{center}
\includegraphics[width=1\columnwidth,angle=0]{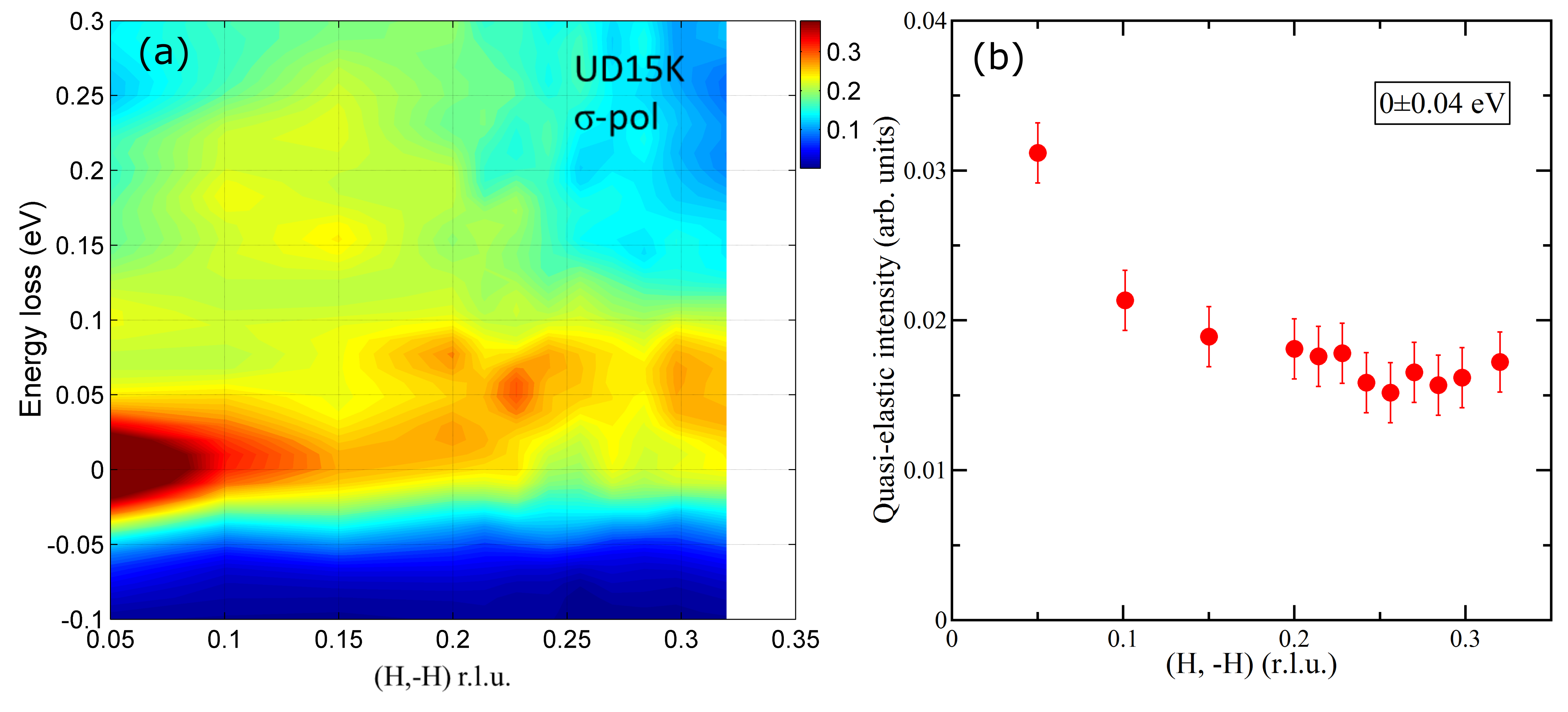}
\end{center}
\caption{(color online) (a) Energy/momentum intensity map of RIXS spectra
along (0,0)-(0.5,-0.5) symmetry direction taken with $\sigma$-polarized
incident light at 20~K for UD15K. (b) Corresponding quasi-elastic
intensity (integrated around 0 eV energy loss over $\pm$0.04 eV range).}
\end{figure}

So far, we have focused on the charge order along (0,0.5) or (0.5,0)
direction. As already noted, the energy-integrated RXS measurement on
Bi2201 gave no CDW signatures along the diagonal direction
\cite{nodalCDW}. However, the checkerboard-like features observed
by STM [\onlinecite{WDWise}] might be induced by two kinds of charge
modulation patterns, either along the Cu-O bond directions or along the
nodal (diagonal) directions. Here we exploited the higher sensitivity of
energy-resolved RIXS to ascertain this issue. As demonstrated above, UD15K
shows a relatively strong CDW signal along (0,0.5) or (0.5,0) direction
with $Q_{\mathrm{CDW}} \simeq 0.26$ rlu, which allows reaching the
($Q_{\mathrm{CDW}}$,-$Q_{\mathrm{CDW}}$) point; on the contrary in YBCO
the hypothetical diagonal point at (0.31,0.31) is out of reach for Cu
$L_3$ RIXS [\onlinecite{GiacomoCDW}]. As shown in Fig.~4(a), along
(0.5,-0.5) direction, the energy/momentum intensity map of RIXS spectra
shows no charge order signal at (0.26,-0.26) rlu around the quasi-elastic
energy region, while there are clear phonon signals present at $\sim$55
meV. The quasi-elastic integrated intensity in Fig.~4(b) does not show any
peak, in good agreement with prior results of RXS measurement on UD15K
[\onlinecite{nodalCDW}]. This confirms that the charge density modulates,
unidirectionally, only along the Cu-O bonds. The
orientation of CDW in cuprates has been discussed recently in a
theoretical work based on the frustrated phase-separation model: along the
diagonal the short-range residual repulsion is stronger than along Cu-O
bonds so that the local effective attraction stablizes the
unidirectional CDW along the [1,0]/[0,1] direction instead of the [1,1] direction
\cite{Grilliarxiv}.

\subsection{Temperature dependence}

\begin{figure}[tbp]
\begin{center}
\includegraphics[width=1\columnwidth,angle=0]{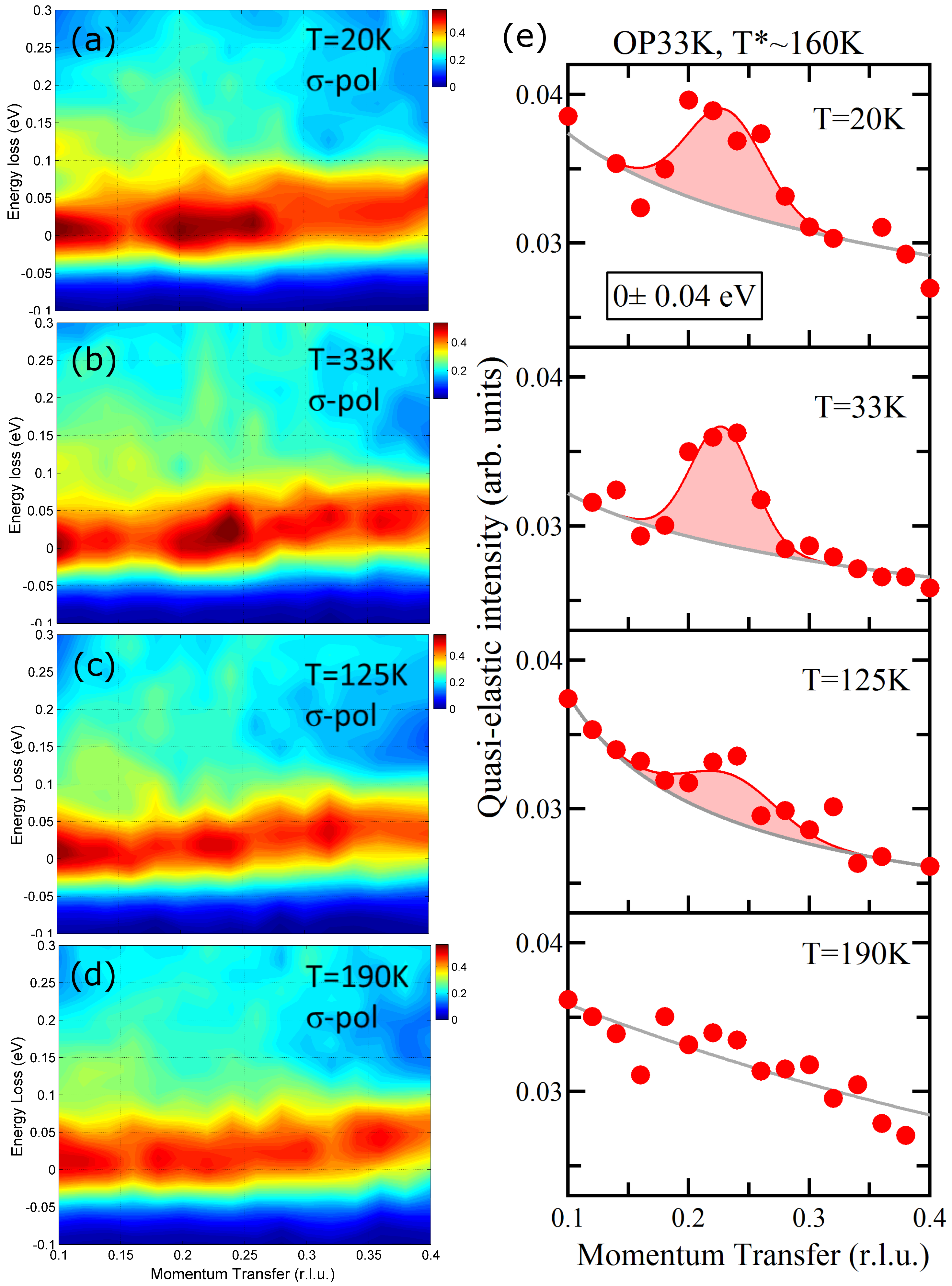}
\end{center}
\caption{(color online) Temperature dependence of charge order across
$T_{c}$$\simeq$33~K and $T^{*}$$\simeq$160~K for OP33K. Energy/momentum
intensity maps of RIXS spectra along (0,0)-(0.5,0) symmetry direction
measured at (a) 20~K, (b) 33~K, (c) 125~K and (d) 190~K with $\sigma$
polarization on OP33K. (e) Corresponding intensity at 0$\pm$0.04 eV for
the quasi-elastic signal. Solid lines are Lorentzian fits to the data with a
power law background. The areas after subtracting the background are
highlighted by the red shading.}
\end{figure}

The temperature dependence of the charge order in UD15K has been reported
in Ref.~\onlinecite{Damascelliscience} with $T_{\mathrm{CDW}} \simeq 180$K
($\sim T^*$). In Fig.~5, we investigated the temperature dependence of the
charge order across $T_{c}$ and $T^{*}$ for OP33K. From previous nuclear
magnetic resonance (NMR) measurements on
Bi$_{2}$Sr$_{2-x}$La$_x$CuO$_{6+\delta}$ [\onlinecite{NMRBi2201}], we know that
the pseudogap temperature of OP33K is $T^* \simeq 160$K. As shown in
Fig.~5(a) and (b), the charge order can been seen clearly at 20K and
becomes sharper at $T_{c} \simeq 33$K, with the width decreasing from
0.07$\pm$0.01 rlu to 0.06$\pm$0.01 rlu, as shown in Fig.~5(e). This behavior of the charge order is
similar to that in YBCO [\onlinecite{GiacomoCDW}], reflecting the competition
between CDW order and superconductivity. Above $T_\mathrm{c}$ the
intensity of the charge order signal progressively decreases
\cite{GiacomoCDW,Damascelliscience}. The charge order peak is still
visible at 125K, below $T^{*}$, but it disappears at 190K, above $T^{*}$.
Since the CDW onset temperature is not a thermodynamic phase boundary and
given the statistical uncertainty in the high temperature RIXS data, we
are not able to determine whether the pseudogap formation precedes or
coincides with the CDW order in OP33K.

\subsection{\textit{L} dependence}

\begin{figure}[tbp]
\begin{center}
\includegraphics[width=0.95\columnwidth,angle=0]{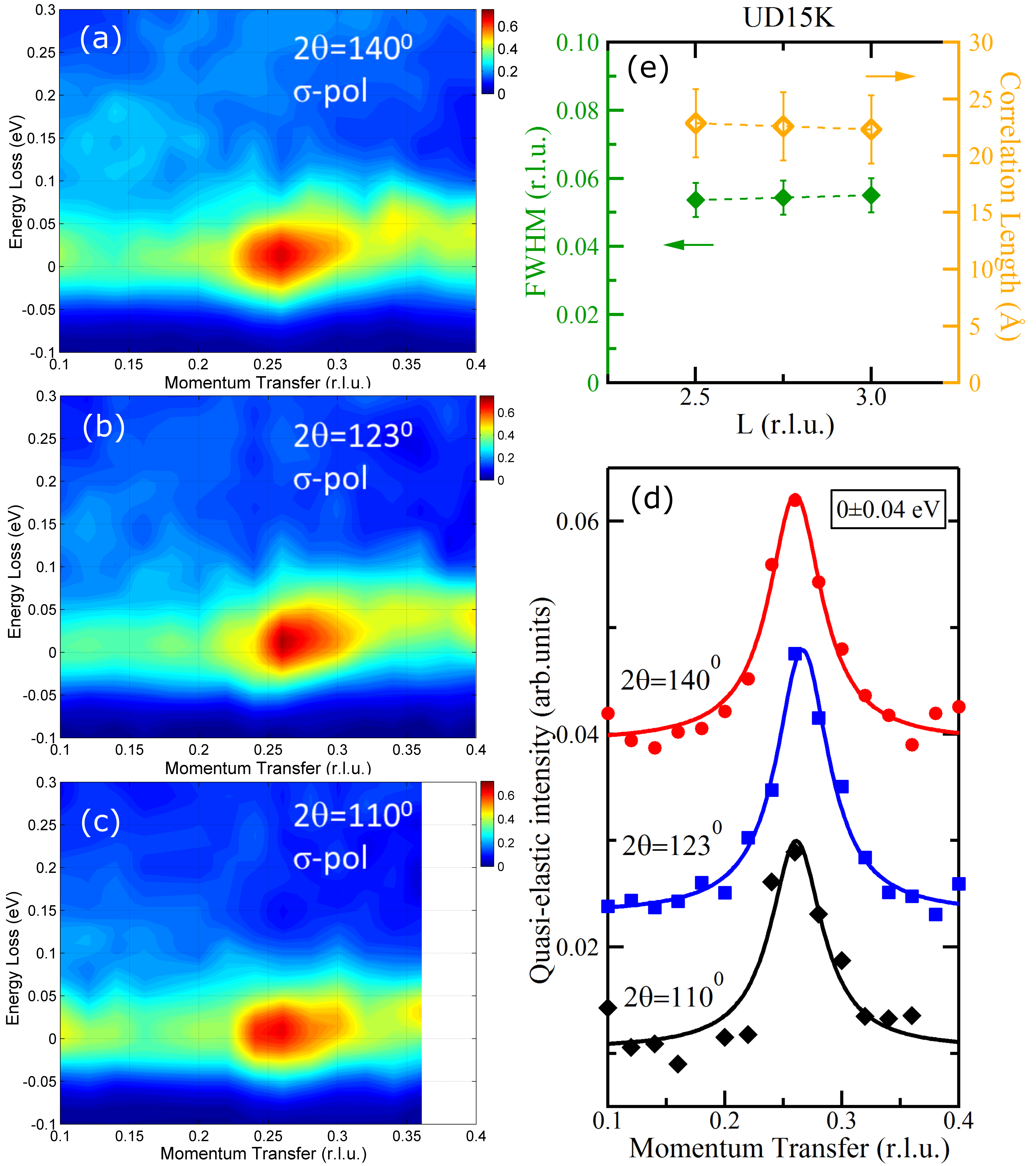}
\end{center}
\caption{(color online) Energy/momentum intensity maps of RIXS spectra
along (0,0)-(0.5,0) symmetry direction measured at (a)
$2\theta=140$ deg, (b) $2\theta=123$ deg and (c)
$2\theta=110$ deg with $\sigma$ polarization for UD15K at 20~K. (d)
Integral intensity at 0$\pm$0.04 eV for the quasi-elastic signal. Data are
shifted vertically for clarity. Solid lines are Lorentzian fits to the data.
(e) \emph{L} dependence of CDW peak FWHM, given in reciprocal lattice
units (rlu) (left) and corresponding correlation length (right) at
20~K. Dashed lines are guides to the eye. }
\end{figure}

\begin{figure*}[htbp]
\begin{center}
\includegraphics[width=2\columnwidth,angle=0]{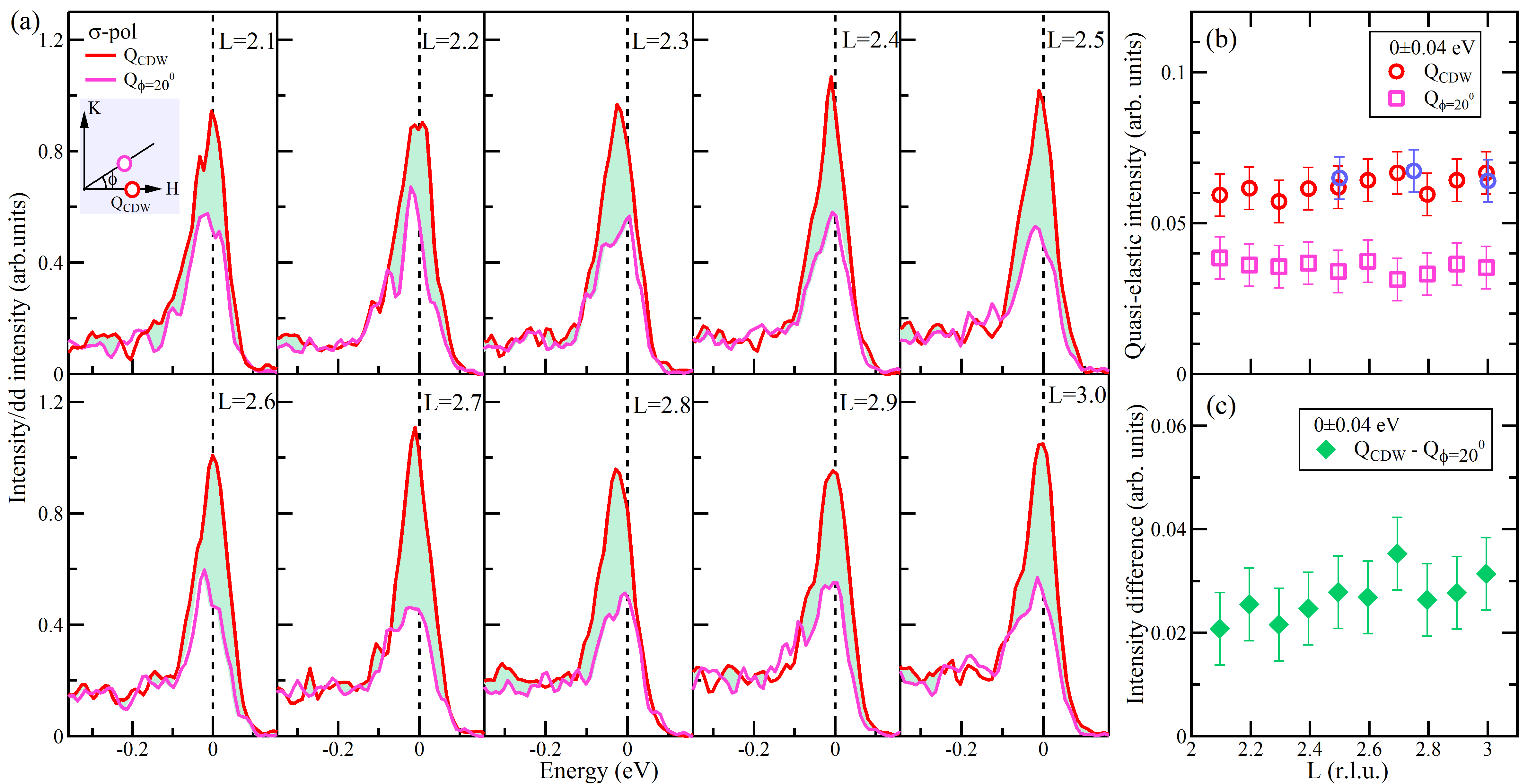}
\end{center}
\caption{(color online) (a) Comparison of RIXS spectra of UD15K at
$Q_{\mathrm{CDW}}$ and at $Q_{\phi}$ with $\phi=20^0$ as defined in the inset,
measured with $\sigma$ polarization from \emph{L}=2.1 to 3.0 rlu. Data
were collected at 20~K. The differences between two spectra at the same
\emph{L} are highlighted by the green shading. Self-absorption correction
has been applied to the intensities. (b) \emph{L} dependence of the
quasi-elastic intensity within 0$\pm$0.04 eV at $Q_{\mathrm{CDW}}$ (hollow red
circles) and at $Q_{\phi=20^0}$ (hollow magenta square). Three data points
from Fig.~6 are also included (hollow blue circles). (c) The intensity
differences between $Q_{\mathrm{CDW}}$ and $Q_{\phi=20^0}$. The error bars
represent the uncertainty in determining the spectral weight. }
\end{figure*}

At zero magnetic field, charge order peaks at half integer values of $L$
in the out-of-plane direction both in YBCO
[\onlinecite{WSLeeHighB},\onlinecite{changNComm}] and LBCO
[\onlinecite{JapaneseLBCO},\onlinecite{Stripexray},\onlinecite{CDWLBCOPRB}].
Here, by exploiting the unique possibility available at ID32 of changing
continuously the scattering angle in RIXS, we investigated the \emph{L}
dependence of CDW in UD15K, to see whether or not there is an intensity
maximum at half integer. In Fig.~6(a-c) we show the energy/momentum
intensity maps measured with three scattering angles ($2\theta =
140^\circ$, 123$^\circ$, 110$^\circ$), corresponding to the charge order
at (0.26, 0, 3), (0.26, 0, 2.75) and (0.26, 0, 2.5). We can see a strong
charge order signals in all three maps with similar quasi-elastic profiles
as shown in Fig.~6(d). By fitting the data we obtained similar FWHMs
($\sim$0.055 rlu) and correlation lengths ($\sim 23$ \AA) for UD15K. We
also performed a finer \emph{L}-scanning as shown in Fig.~7. For
simplicity, instead of taking a full map at each $L$, we took a couple of
RIXS spectra for each $L$, at \textbf{$Q_{\mathrm{CDW}}$} and at
$|Q_{\phi}|$, with $\phi =20$ deg in-plane rotation as defined in the
inset and $|Q_{\phi}|=|Q_{\mathrm{CDW}}|$. The \emph{L} values are ranging
from 2.1 to 3.0 rlu. The difference between two spectra at the same
\emph{L} is marked by the green area, indicating the charge order
intensity. In Fig. 7(b) we display the integral quasi-elastic intensities
at \textbf{$Q_{\mathrm{CDW}}$} and at \textbf{$Q_{\phi}$}: neither of them
display a maximum within error bar, not even at half integer $L$. In
Fig.~7(c) we plot the intensity difference between the two momenta and
again it shows no maximum at half integer nor at other values. Clearly the
CDW intensity is flat across an entire Brillouin zone, demonstrating the
complete absence of correlation along the \emph{c}-axis.

\section{Discussion}
\subsection{Relation of charge order to antiferromagnetic and pseudogap state}

We have found that charge order persists in Bi2201 up to optimal doping,
consistently with LBCO [\onlinecite{CDWLBCOPRB}], Bi2212
[\onlinecite{HashimotoBi2212}]
and YBCO [\onlinecite{TaconYBCOPRB}]. Our recent RIXS study on antiferromagnetic
Bi2201 ($p=0.03$) showed no charge order signals \cite{AFcuprate}. A
recent STM experiment has reported a checkerboard-like charge order with
wave vector $Q_{\mathrm{CDW}}$$\simeq$0.25 rlu in lightly doped Bi2201
($p=0.07$, insulator) \cite{CaiSTM}, demonstrating the charge order is the
first electronic ordered state that emerges by doping the parent compound.
The same work also confirmed the absence of checkerboard-like pattern at very
low doping \emph{p}=0.03 [\onlinecite{CaiSTM}], in agreement with our RIXS
results. The evidence of charge order in the AF insulating regime is
incompatible with the ``Fermi arc nesting" scenario which correlates the
$Q_{\mathrm{CDW}}$ to the distance between the Fermi-arc tips
\cite{Damascelliscience}, because in single layer Bi2201 the Fermi surface
is fully gapped below the antiferromagnetic critical point $p=0.1$
[\onlinecite{PengNaturecomm}]. Note that it has been proposed that the
emergence of the checkerboard structure is a consequence of the proximity
to the quantum critical point \cite{PGQCPNP,PGQCPPRB}.

In the SC regime above the critical doping \emph{p}=0.1, the charge order
vector and its correlation length decrease with doping. As summarized in
Fig.~3, the CDW order in Bi2201 has been experimentally detected for
$p_{c_1}$$\leqslant$\emph{p}$\leqslant$$p_{c_2}$ with
$p_{c_1}$$\backsimeq$0.07 and $p_{c_2}$$\backsimeq$0.16. We notice that
there is a non-negligible discrepancy between charge order vectors
determined from STM and RXS. The latter gives in generally larger CDW
vectors. The difference increases with doping and reaches 35$\%$ in the
optimal doping, which is out of the error tolerance. This may reflect the
bulk (x-ray scattering) and surface (STM) dichotomy of charge order. It is
known that the carrier concentration is different at the surface and in
the bulk, and that the difference grows with doping. The CDW order arising
from the charge modulations can reflect this dichotomy: indeed at low
doping $p \simeq 0.11$ the $Q_{\mathrm{CDW}}$ vectors are similar between
RXS and STM results, while they separate into two trends and departs
further with doping. A recent STM work, by utilizing the phase-resolved
electronic structure visualization, has revealed a surprising doping
independent locking of the local CDW wavevector at 0.25 rlu throughout the
underdoped phase diagram of (bi-layer) Bi2212 [\onlinecite{JCDavisarxiv}]. While
the generality of this lattice-commensurability CDW in other cuprates, and
the supposed correlation with x-ray scattering results through phase slips
between different short-ranged correlated domains remain to be explored,
at least the conventional CDW amplitudes probed by x-ray and STM look
different.

If compared to other cuprates, charge order in Bi2201 is rather
short-ranged, with a real-space correlation length between 17 {\AA} and 23 \AA,
similar to $\sim 24$ {\AA} in Bi2212 [\onlinecite{HashimotoBi2212}], but
shorter than $20 - 70$ {\AA} in YBCO [\onlinecite{TaconYBCOPRB}] and
$150 - 250$ {\AA} in LBCO
[\onlinecite{Stripexray},\onlinecite{CDWLBCOPRB}]. Probably lager disorder
(e.g. chemical inhomogeneity) plays a bigger role in Bi2201 and Bi2212
[\onlinecite{EisakiPRB}]. There is still no report of charge order in
overdoped regime of cuprates. For Bi2201 the pseudogap state extends to the
heavily overdoped regime, which is well defined by the NMR measurements
\cite{NMRBi2201} (Fig.~1(a)). Therefore the end point of pseudogap
state does not coincide with the end point of charge order in Bi2201. Also
in YBCO the high magnetic field Hall coefficient measurements showed that
the Fermi surface reconstruction by charge order ended at the optimal
doping \emph{p}=0.16 [\onlinecite{psudogapCP}], which is distinctly lower
than the pseudogap critical point \emph{p}=0.19 [\onlinecite{YBCOPGend}].
As Keimer et al. [\onlinecite{KeimerNature}] have already pointed out, the
pseudogap is characterized by several competing ordering tendencies and it
would not be surprising that the critical points between pseudogap and
charge order do not coincide, although the opposite is not excluded too.
Moreover, which termination point (as determined by Fermi surface
reconstruction \cite{psudogapCP,HoffmanSTMFS},
symmetry breaking \cite{JCDavisSTM}, divergent effective mass
\cite{QPmassenhancement}, etc.) actually relates to the QCP is
hotly debated, and there is a need for further experimental and
theoretical investigations.

\subsection{Correlation of charge order along the \emph{c}-direction}

Besides the in-plane components, x-ray studies of the charge order in YBCO
and LBCO have reported an additional correlation along the
\emph{c}-direction
\cite{JapaneseLBCO,Stripexray,CDWLBCOPRB,changCDW,WSLeeHighB,changNComm}.
This depends on the relative (from plane to plane) phase of the CDW
modulations along the \emph{c}-direction. In zero magnetic field, x-ray
diffraction of the ionic displacements in YBCO [\onlinecite{CDWantiphase}]
revealed a weak anti-phase correlation in neighboring bilayers, which
results in the CDW peaking at half-integral values of \emph{L}. This finite
\emph{c}-axis coherence of the CDW is rather short-range, with a length of
$\backsim 9 \pm3$ \AA (i.e. approximately the distance between two
adjacent bi-layers) \cite{CDWantiphase}. By applying high magnetic
fields (typically $B>15$ T), the \emph{c}-direction behavior of the CDW
evolves differently along the \emph{a}- and \emph{b}-directions: the
correlation simply becomes stronger in \emph{a}-direction while a new peak
appears at \emph{L}=1 along the \emph{b}-direction with increasing
correlation length $\sim 4c$ [\onlinecite{changNComm}]. This means the
CDWs propagating along the \emph{a}-axis keep anti-phase correlation
between neighboring bi-layers while those propagating along the
\emph{b}-axis lock their phase with neighboring bi-layers. In LBCO the
charge stipe order also exhibits broad maxima at half-integer \emph{L},
indicative of a twofold periodicity along the \emph{c} axis
\cite{Stripexray,CDWLBCOPRB,JapaneseLBCO}. The reason is that in
adjacent planes within one unit cell the stripes align in orthogonal
directions arising from the tilting pattern of the CuO$_6$ octahedra. In
addition, the charge order is offset by $2a$ between successive unit
cells, presumably to minimize Coulomb repulsion, resulting in an antiphase
relationship between next-nearest-neighbor CuO$_2$ planes
\cite{Stripexray,CDWLBCOPRB,JapaneseLBCO}. This out-of-plane
correlation is very short-ranged $\sim 5-10$ $\AA$ ($< \emph{c}$) and can be
enhanced by a magnetic field \cite{mageticLBCO}.

On the other hand, our $c$-direction study of CDW in Bi2201 does not show
any peak, indicating there is no phase correlation of CDW along the
\emph{c}-direction. The underlying reason can be the following: the
distance between the adjacent Cu-O planes in Bi2201 is $\sim 12.2$ {\AA}
within one unit cell, and neighboring planes are offset by (0.5,0.5): CDW
correlation is discouraged both by distance and crystalline mismatch,
resulting in random phases along the \emph{c}-direction. There is no high
magnetic field measurement on Bi2201 yet, but the 2D CDW is to be
expected, since the coupling between the two Cu$O_2$ planes within one
unit cell is rather weak, let alone the coupling between two unit cells.
For double layer Bi2212, we infer that the charge order also has no phase
correlation along the \emph{c}-direction, because the distance between two
bilayers is $\sim$12.3 \AA, much larger than in YBCO ($\sim$8.5 \AA),
which maybe also discouraged by the (0.5,0.5) offset.

\section{Conclusions}

We have directly observed a bulk, incommensurate charge order in UD15K and
OP33K, demonstrating that the short-range charge density modulations
persist up to optimal doping in Bi2201. Both the CDW intensity and
correlation length decrease with doping. In addition, the CDW vector
decreases with doping, showing a bulk/surface dichotomy from RXS and STM
measurements. The doping range of charge order in Bi2201, $p_{c_1}
\leqslant p \leqslant p_{c_2}$ with $p_{c_1}$$\backsimeq$0.07 and
$p_{c_2}$$\backsimeq$0.16, suggests the critical points of charge order
are different from those of AF, SC and pseudogap. Thus, charge order
appears to be a separate phenomenon that coexists with AF, SC and
pseudogap. Temperature measurements have demonstrated that it competes
with superconductivity and the signal disappears across $T^*$ due to
fluctuations. Whether and how it relates to QCP for the mechanism of
high-$T_\mathrm{c}$ superconductivity requires future experimental and
theoretical research. Furthermore, we confirmed there is no charge order
along the diagonal direction, suggesting the CDW propagates only along the
Cu-O bond direction. This fact is also compatible with the observed
\emph{L}-independence of CDW, indicating there is no phase correlations
along the \emph{c}-direction at variance to YBCO and LBCO, and hinting at
a perfectly two-dimensional charge ordering in single layer Bi2201.

\begin{acknowledgments}
The experimental data were collected at the beam line ID32 of the European
Synchrotron (ESRF) in Grenoble (F) using the ERIXS spectrometer designed
jointly by the ESRF and Politecnico di Milano. This work was supported by
MIUR Italian Ministry for Research through project PIK Polarix. We thank
Marc-Henri Julien for his help with the characterization of some samples. We
gratefully acknowledge the support of all the staff of the ID32 beam line
of the ESRF, in particular Flora Yakhou-Harris, Andrea Fondacaro and Andrea
Amorese.
\end{acknowledgments}


\begin{thebibliography}{0}%
\makeatletter
\providecommand \@ifxundefined [1]{%
 \@ifx{#1\undefined}
}%
\providecommand \@ifnum [1]{%
 \ifnum #1\expandafter \@firstoftwo
 \else \expandafter \@secondoftwo
 \fi
}%
\providecommand \@ifx [1]{%
 \ifx #1\expandafter \@firstoftwo
 \else \expandafter \@secondoftwo
 \fi
}%
\providecommand \natexlab [1]{#1}%
\providecommand \enquote  [1]{``#1''}%
\providecommand \bibnamefont  [1]{#1}%
\providecommand \bibfnamefont [1]{#1}%
\providecommand \citenamefont [1]{#1}%
\providecommand \href@noop [0]{\@secondoftwo}%
\providecommand \href [0]{\begingroup \@sanitize@url \@href}%
\providecommand \@href[1]{\@@startlink{#1}\@@href}%
\providecommand \@@href[1]{\endgroup#1\@@endlink}%
\providecommand \@sanitize@url [0]{\catcode `\\12\catcode `\$12\catcode
  `\&12\catcode `\#12\catcode `\^12\catcode `\_12\catcode `\%12\relax}%
\providecommand \@@startlink[1]{}%
\providecommand \@@endlink[0]{}%
\providecommand \url  [0]{\begingroup\@sanitize@url \@url }%
\providecommand \@url [1]{\endgroup\@href {#1}{\urlprefix }}%
\providecommand \urlprefix  [0]{URL }%
\providecommand \Eprint [0]{\href }%
\providecommand \doibase [0]{http://dx.doi.org/}%
\providecommand \selectlanguage [0]{\@gobble}%
\providecommand \bibinfo  [0]{\@secondoftwo}%
\providecommand \bibfield  [0]{\@secondoftwo}%
\providecommand \translation [1]{[#1]}%
\providecommand \BibitemOpen [0]{}%
\providecommand \bibitemStop [0]{}%
\providecommand \bibitemNoStop [0]{.\EOS\space}%
\providecommand \EOS [0]{\spacefactor3000\relax}%
\providecommand \BibitemShut  [1]{\csname bibitem#1\endcsname}%
\let\auto@bib@innerbib\@empty
\end{thebibliography}%


\begin{thebibliography}{99}

\bibitem{HighTcuprate} J. G. Bednorz and K. A. M\"{u}ller, Z. Phys. B \textbf{64},
    189-193 (1986).
\bibitem{PALeeReview} P. A. Lee, N. Nagaosa, and X. G. Wen, Rev. Mod.
    Phys. \textbf{78}, 17 (2006).
\bibitem{PGreview} T. Timusk and B. Statt, Rep. Prog. Phys. \textbf{62}, 61 (1999).
\bibitem{TranquadaStripe} J. M. Tranquada, B. J. Sternlieb, J. D. Axe, Y.
    Nakamura and S. Uchida, Nature (London) \textbf{375}, 561 (1995).
\bibitem{FujitaStripe} M. Fujita, H. Goka, K. Yamada, and M. Matsuda,
    Phys. Rev. Lett. \textbf{88}, 167008 (2002).
\bibitem{JapaneseLBCO} H. Kimura, H. Goka, M. Fujita, Y. Noda, K. Yamada,
    and N. Ikeda, Phys. Rev. B \textbf{67}, 140503 (2003).
\bibitem{Stripexray} P. Abbamonte, A. Rusydi, S. Smadici, G. D. Gu, G. A.
    Sawatzky, and D. L. Feng, Nat. Phys. \textbf{1}, 155 (2005).
\bibitem{CDWLBCOPRB} M. H\"{u}cker, M. v. Zimmermann, G. D. Gu, Z. J. Xu, J.
    S. Wen, Guangyong Xu, H. J. Kang, A. Zheludev and J. M. Tranquada,
    Phys. Rev. B \textbf{83}, 104506 (2011).
\bibitem{JulienNMR} T. Wu, H. Mayaffre, S. Kr\"amer, M. Horvati\'{c}, C.
    Berthier, W. N. Hardy, R. X. Liang, D. A. Bonn, and M.-H. Julien,
    Nature (London) \textbf{477}, 191 (2011).
\bibitem{GiacomoCDW} G. Ghiringhelli, M. Le Tacon, M. Minola, S.
    Blanco-Canosa, C. Mazzoli, N. B. Brookes, G. M. De Luca, A. Fra\~{n}o, D.
    G. Hawthorn, F. He, T. Loew, M.M. Sala, D. C. Peets, M. Salluzzo, E.
    Schierle, R. Sutarto, G. A. Sawatzky, E. Weschke, B. Keimer, and L.
    Braicovich, Science \textbf{337}, 821 (2012).
\bibitem{RXSCDW} A. J. Achkar, R. Sutarto, X. Mao, F. He, A. Fra\~{n}o, S.
    Blanco-Canosa, M. Le Tacon, G. Ghiringhelli, L. Braicovich, M. Minola, M.
    Moretti Sala, C. Mazzoli, R. Liang, D. A. Bonn,W. N. Hardy, B. Keimer, G.
    A. Sawatzky, and D. G. Hawthorn, Phys. Rev. Lett. \textbf{109}, 167001 (2012).
\bibitem{changCDW} J. Chang, E. Blackburn, A. T. Holmes, N. B.
    Christensen, J. Larsen, J. Mesot, R. Liang, D. A. Bonn, W. N. Hardy,
    A. Watenphul, M. v. Zimmermann, E. M. Forgan, and S. M. Hayden, Nat.
    Phys. \textbf{8}, 871 (2012).
\bibitem{diffractionCDW} E. Blackburn, J. Chang, M. H\"{u}cker, A. T. Holmes,
    N. B. Christensen, R. Liang, D. A. Bonn, W. N. Hardy, U. R\"{u}tt, O.
    Gutowski, M. v. Zimmermann, E. M. Forgan, and S. M. Hayden, Phys. Rev.
    Lett. \textbf{110}, 137004 (2013).
\bibitem{Damascelliscience} R. Comin, A. Fra\~{n}o, M. M. Yee, Y. Yoshida, H.
    Eisaki, E. Schierle, E. Weschke, R. Sutarto, F. He, A. Soumyanarayanan, Y.
    He, M. Le Tacon, I. Elfimov, J. E. Hoffman, G. Sawatzky, B. Keimer, and A.
    Damascelli, Science \textbf{343}, 390 (2014).
\bibitem{STMBi2212} E. H. da Silva Neto, P. Aynajian, A. Fra\~{n}o, R. Comin,
    E. Schierle, E. Weschke, A. Gyenis, J. S. Wen, J. Schneeloch, Z. J.
    Xu, S. Ono, G. D. Gu, M. Le Tacon, and Ali Yazdani, Science \textbf{343}, 393
    (2014).
\bibitem{HashimotoBi2212} M. Hashimoto, G. Ghiringhelli, W.-S. Lee, G.
    Dellea, A. Amorese, C. Mazzoli, K. Kummer, N. B. Brookes, B. Moritz,
    Y. Yoshida, H. Eisaki, Z. Hussain, T. P. Devereaux, Z.-X. Shen, and L.
    Braicovich, Phys. Rev. B \textbf{89}, 220511(R) (2014).
\bibitem{HBCOCDW} W. Tabis, Y. Li, M. Le Tacon, L. Braicovich, A.
    Kreyssig, M. Minola, G. Dellea, E. Weschke, M. J. Veit, M.
    Ramazanoglu, A. I. Goldman, T. Schmitt, G. Ghiringhelli, N. Bari\v{s}i\'{c},
    M. K. Chan, C. J. Dorow, G. Yu, X. Zhao, B. Keimer, and M. Greven,
    Nat. Commun. \textbf{5}, 5875 (2014).
\bibitem{TaconYBCOPRB} S. Blanco-Canosa, A. Fra\~{n}o, E. Schierle, J. Porras,
    T. Loew, M. Minola, M. Bluschke, E. Weschke, B. Keimer, M. Le Tacon,
    Phys. Rev. B \textbf{90}, 054513 (2014).
\bibitem{CominNBCO} E. H. da Silva Neto, R. Comin, F. He, R. Sutarto, Y.
    Jiang, R. L. Greene, G. A. Sawatzky, and A. Damascelli, Science,
    \textbf{347}, 282 (2015).
\bibitem{EduardoSA} E. H. da Silva Neto, B. Q. Yu, M. Minola, R. Sutarto,
    E. Schierle, F. Boschini, M. Zonno, M. Bluschke, J. Higgins, Y. M. Li,
    G. C. Yu, E. Weschke, F. Z. He, M. Le Tacon, R. L. Greene, M. Greven,
    G. A. Sawatzky, B. Keimer and A. Damascelli, Science Advances \textbf{2},
    1600782 (2016).
\bibitem{KeimerNature} B. Keimer, S. A. Kivelson, M. R. Norman, S. Uchida
    and J. Zaanen, Nature \textbf{518},179 (2015).
\bibitem{quantumoscillation1} N. Doiron-Leyraud, C. Proust, D. LeBoeuf, J.
    Levallois, J. B. Bonnemaison, R. X. Liang, D. A. Bonn, W. N. Hardy,
    and L. Taillefer, Nature (London) \textbf{447}, 565 (2007).
\bibitem{quantumoscillation2} S. E. Sebastian, N. Harrison, E. Palm, T. P.
    Murphy, C. H. Mielke, R. X. Liang, D. A. Bonn, W. N. Hardy, and G. G.
    Lonzarich, Nature (London) \textbf{454}, 200 (2008).
\bibitem{PengPRB} Y. Y. Peng, M. Hashimoto, M. Moretti Sala, A. Amorese,
    N. B. Brookes, G. Dellea, W.-S. Lee, M. Minola, T. Schmitt, Y.
    Yoshida, K.-J. Zhou, H. Eisaki, T. P. Devereaux, Z.-X.
    Shen, L. Braicovich, and G. Ghiringhelli, Phys. Rev. B \textbf{92}, 064517
    (2015).
\bibitem{Hashimotonphys} M. Hashimoto, R.-H. He, K. Tanaka, J.-P. Testaud,
    W. Meevasana, R. G. Moore, D. H. Lu, H. Yao, Y. Yoshida, H. Eisaki, T.
    P. Devereaux, Z. Hussain and Z.-X. Shen, Nat. Phys. \textbf{6}, 414 (2010).
\bibitem{Hescience} R.-H. He, M. Hashimoto, H. Karapetyan, J. D. Koralek,
    J. P. Hinton, J. P. Testaud, V. Nathan, Y. Yoshida, Hong Yao, K.
    Tanaka, W. Meevasana, R. G. Moore, D. H. Lu, S.-K. Mo, M. Ishikado, H.
    Eisaki, Z. Hussain, T. P. Devereaux, S. A. Kivelson, J. Orenstein, A.
    Kapitulnik, and Z.-X. Shen, Science \textbf{331}, 1579 (2011).
\bibitem{WDWise} W. D. Wise, M. C. Boyer, Kamalesh Chatterjee, Takeshi
    Kondo, T. Takeuchi, H. Ikuta, Yayu Wang and E. W. Hudson, Nat. Phys.
    \textbf{4}, 696-699 (2008).
\bibitem{nodalCDW} R. Comin, R. Sutarto, F. He, E. H. da Silva Neto, L.
    Chauviere, A. Fra\~{n}o, R. Liang, W. N. Hardy, D. A. Bonn, Y. Yoshida, H.
    Eisaki, A. J. Achkar, D. G. Hawthorn, B. Keimer, G. A. Sawatzky and A.
    Damascelli, Nat. Materials \textbf{14}, 796 (2015).
\bibitem{CDWantiphase} E. M. Forgan, E. Blackburn, A. T. Holmes, A. K. R.
    Briffa, J. Chang, L. Bouchenoire, S. D. Brown, R. X. Liang, D. Bonn,
    W. N. Hardy, N. B. Christensen, M.v. Zimmermann, M. H\"{u}cker and S. M.
    Hayden, Nat. Commun. \textbf{6}, 10064 (2015).
\bibitem{NMRBi2201} S. J. Kawasaki, C. T. Lin, P. L. Kuhns, A. P. Reyes,
    and G.-Q. Zheng, Phys. Rev. Lett. \textbf{105}, 137002 (2010).
\bibitem{meng} J. Q. Meng, G. D. Liu, W. T. Zhang, L. Zhao, H. Y. Liu, W.
    Lu, X. L. Dong and X. J. Zhou, Supercond. Sci. Technol. \textbf{22},
    045010(2009).
\bibitem{ZChen} Z. Chen, Y. Y. Peng, Z. Wang, Y. J. Song, J. Q. Meng, X.
    J. Zhou and J. Q. Li, Supercond. Sci. Technol. \textbf{26}, 055010 (2013).
\bibitem{Luciomagnon} L. Braicovich, M. Moretti Sala, L. J. P. Ament, V.
    Bisogni, M. Minola, G. Balestrino, D. Di Castro, G. M. De Luca, M.
    Salluzzo, G. Ghiringhelli, and J. van den Brink, Phys. Rev. B
    \textbf{81}, 174533 (2010).
\bibitem{CaiSTM} P. Cai, W. Ruan, Y. Y. Peng, C. Ye, X. T. Li, Z. Q. Hao,
    X. J. Zhou, D.-H. Lee and Y. Y. Wang, Nat. Phys. DOI:
    10.1038/NPHYS3840 (2016).
\bibitem{Grilliarxiv} S. Caprara, C. Di Castro, G. Seibold and M. Grilli,
    arXiv:1604.07852v1 (2016).
\bibitem{WSLeeHighB} S. Gerber, H. Jang, H. Nojiri, S. Matsuzawa, H.
    Yasumura, D. A. Bonn, R. Liang, W. N. Hardy, Z. Islam, A. Mehta, S.
    Song, M. Sikorski, D. Stefanescu, Y. Feng, S. A. Kivelson, T. P.
    Devereaux, Z.-X. Shen, C.-C. Kao, W.-S. Lee, D. Zhu, J.-S. Lee,
    Science \textbf{350}, 949 (2015).
\bibitem{changNComm} J. Chang, E. Blackburn, O. Ivashko, A. T. Holmes, N. B.
    Christensen, M. H\"{u}cker, R. Liang, D. A. Bonn, W. N. Hardy, U. R\"{u}tt,
    M. v. Zimmermann, E. M. Forgan and S. M. Hayden,  Nat. Commun. \textbf{7},
    11494 (2016).
\bibitem{AFcuprate} Y. Y. Peng, G. Dellea, M. Minola, M. Conni, A.
    Amorese, D. Di Castro, G. M. De Luca, K. Kummer, M. Salluzzo, X. Sun,
    X. J. Zhou, G. Balestrino, M. Le Tacon, B. Keimer, L. Braicovich, N.
    B. Brookes and G. Ghiringhelli, arXiv:1609.05405 (2016).
\bibitem{PengNaturecomm} Y. Y. Peng, J. Q. Meng, D. X. Mou, J. F. He, L.
    Zhao, Y. Wu, G. D. Liu, X. L. Dong, S. L. He, J. Zhang, X. Y. Wang, Q.
    J. Peng, Z. M. Wang, S. J. Zhang, F. Yang, C. T. Chen, Z. Y. Xu, T. K.
    Lee and X. J. Zhou, Nat. Commun. \textbf{4}, 2459 (2013).
\bibitem{PGQCPNP} K. B. Efetov, H. Meier and C. P\'{e}pin, Nat. Phys. \textbf{9},
    442-446 (2013).
\bibitem{PGQCPPRB} H. Meier, C. P\'{e}pin, M. Einenkel, and K. B. Efetov,
    Phys. Rev. B \textbf{89}, 195115 (2014).
\bibitem{JCDavisarxiv} A. Mesaros, K. Fujita, S. D. Edkins, M. H.
    Hamidian, H. Eisaki, S. Uchida, J. C. S\'{e}amus Davis, M. J. Lawler and
    Eun-Ah Kim, arXiv:1608.06180 (2016).
\bibitem{EisakiPRB} H. Eisaki, N. Kaneko, D. L. Feng, A. Damascelli, P. K.
    Mang, K. M. Shen, Z.-X. Shen, and M. Greven, Phys. Rev. B \textbf{69}, 064512
    (2004).
\bibitem{psudogapCP} S. Badoux, W. Tabis, F. Lalibert\'{e}, G. Grissonnanche,
    B. Vignolle, D. Vignolles, J. B\'{e}ard, D. A. Bonn, W. N. Hardy, R.
    Liang, N. Doiron-Leyraud, L. Taillefer and C. Proust, Nature \textbf{531},
    210-214 (2016).
\bibitem{YBCOPGend} J. L. Tallona and J.W. Loramb, Physica C \textbf{349}, 53-68
    (2001).
\bibitem{HoffmanSTMFS} Y. He, Y. Yin, M. Zech, A. Soumyanarayanan, M. M.
    Yee, T. Williams, M. C. Boyer, K. Chatterjee, W. D. Wise, I.
    Zeljkovic, T. Kondo, T. Takeuchi, H. Ikuta, P. Mistark, R. S.
    Markiewicz, A. Bansil, S. Sachdev, E. W. Hudson, J. E. Hoffman,
    Science \textbf{344}, 608 (2014).
\bibitem{JCDavisSTM} K. Fujita, C. K. Kim, I. Lee, J. Lee, M. H. Hamidian,
    I. A. Firmo, S. Mukhopadhyay, H. Eisaki, S. Uchida, M. J. Lawler,
    E.-A. Kim, J. C. Davis, Science \textbf{344}, 612 (2014).
\bibitem{QPmassenhancement} B. J. Ramshaw, S. E. Sebastian, R. D.
    McDonald, J. Day, B. S. Tan, Z. Zhu, J. B. Betts, R. Liang, D. A.
    Bonn, W. N. Hardy, and N. Harrison, Science \textbf{348}, 317 (2015).
\bibitem{mageticLBCO} M. H\"{u}cker, M. v. Zimmermann, Z. J. Xu, J. S. Wen,
    G. D. Gu, and J. M. Tranquada, Phys. Rev. B \textbf{87}, 014501 (2013).





\end{thebibliography}
\end{document}